\documentclass[useAMS,usenatbib]{mn2e}

\usepackage{graphicx}
\usepackage{txfonts}
\usepackage[breaklinks=true]{hyperref}
\usepackage{longtable}
\usepackage{subfig}
\usepackage{float}
\usepackage{booktabs}
\usepackage{natbib}
\usepackage{ amssymb }
\newcommand{\gsim}{\lower.7ex\hbox{$\;\stackrel{\textstyle>}{\sim}\;$}}

\def\arcsec{\mbox{arcsec}}%
\def\pks{PBC\,J2333.9-2343}
\def\kms{km\,s$^{-1}$}

\def\araa{ARA\&A}%
\def\apj{ApJ}%
\def\apjl{ApJ}%
\def\apjs{ApJS}%
\def\aap{A\&A}%
\def\mnras{MNRAS}%
\def\pasp{PASP}%
\def\ssr{Space~Sci.~Rev.}%
\def\nat{Nature}%

\title[BLR and outflow in PBC\,J2333.9-2343]{Variable broad lines and outflow in the weak blazar PBC\,J2333.9-2343}
\author[Hern\'{a}ndez-Garc\'{i}a et al.]{Hern\'{a}ndez-Garc\'{i}a, L.$^{1,2}$\thanks{E-mail:
lorena.hernandez@uv.cl}, Vietri, G.$^{3,4,5}$, Panessa, F.$^{2}$, Piconcelli, E.$^{3}$, Chavushyan, V.$^{6}$, \newauthor Jim\'enez-Andrade, E.F.$^{7,8}$, Bassani, L.$^{9}$, Bazzano, A.$^{2}$, Cazzoli, S.$^{10}$, Malizia, A.$^{9}$, \newauthor Masetti, N.$^{9,11}$, Monaco, L.$^{11}$, Povi\'c, M.$^{12,10}$, Saviane, I.$^{13}$, Ubertini, P.$^{2}$\\
$^{1}$Instituto de F\'isica y Astronom\'ia, Facultad de Ciencias, Universidad de Valpara\'iso, Gran Breta\~{n}a 1111, Playa Ancha, Valpara\'iso, Chile\\
             $^{2}$INAF - Istituto di Astrofisica e Planetologia Spaziali di Roma (IAPS-INAF), Via del Fosso del Cavaliere 100, 00133 Roma, Italy\\
             $^{3}$INAF - Osservatorio Astronomico di Roma, via Frascati 33, 00078 Monte Porzio Catone, Italy\\
             $^{4}$Excellence Cluster Universe, Technische Universit\"{a}t M\"{u}nchen, Boltzmannstr. 2, D-85748, Garching, Germany \\
             $^{5}$European Southern Observatory, Karl-Schwarzschild-Str. 2, 85748 Garching b. M\"{u}nchen, Germany              \\
             $^{6}$Instituto Nacional de Astrof\'{i}sica, \'{O}ptica y Electr\'{o}nica, Apartado Postal 51-216, 72000 Puebla, M\'{e}xico\\
             $^{7}$Argelander Institute for Astronomy, University of Bonn, Auf dem H\"{u}gel 71, D-53121 Bonn, Germany\\
			 $^{8}$International Max Planck Research School of Astronomy and Astrophysics at the Universities of Bonn and Cologne \\
			 $^{9}$INAF - Osservatorio di Astrofisica e Scienza dello Spazio, via Gobetti 93/3, I-40129 Bologna, Italy \\
			$^{10}$Instituto de Astrof\'{i}sica de Andaluc\'{i}a, CSIC, Glorieta de la Astronom\'{i}a, s/n, 18008 Granada, Spain \\
			 $^{11}$Departamento de Ciencias F\'{i}sicas, Universidad Andr\'{e}s Bello, Fern\'{a}ndez Concha 700, Las Condes, Santiago, Chile \\
			 $^{12}$Ethiopian Space Science and Technology Institute (ESSTI), Entoto Observatory and Research Center (EORC), Addis Ababa, Ethiopia \\
			 $^{13}$European Southern Observatory, Alonso de Cordova 3107, Santiago, Chile \\
             }

\begin{document}

\date{Draft:\today}

\pagerange{\pageref{firstpage}--\pageref{lastpage}} \pubyear{2018}

\maketitle

\label{firstpage}

\begin{abstract}
PBC\,J2333.9-2343 is a peculiar active nucleus with two giant radio lobes and a weak blazar-like nucleus at their center. 
In the present work we show new optical, UV, and X-ray data taken from the San Pedro M\'artir telescope, the New Technology Telescope, NTT/EFOSC2, and the \emph{Swift}/XRT satellite.
The source is highly variable at all frequencies, in particular the strongest variations are found in the broad H$\alpha$ component with a flux increase of 61$\pm$4\ per cent between 2009 and 2016, following the X-ray flux increase of 62$\pm$6\ per cent between 2010 and 2016. We also detected a broad H$\beta$ component in 2016, 
making the optical classification change from type 1.9 to type 1.8 in one year. We have also detected a broad component of the [OIII]$\lambda$5007 line, which is blue-shifted and of high velocity, suggesting an origin from a highly disturbed medium, possibly an outflow.  
The line flux variability and broad widths are indicative of a jet that is, at least in part, responsible for the ionization of the BLR and NLR. 
\end{abstract}


\section{\label{intro}Introduction}

PBC\,2333.9-2343 \citep[z=0.0475,][]{parisi2012}, is a very peculiar active galactic nucleus (AGN), being a giant radio galaxy but hosting in its center a blazar-like nucleus. 
The NVSS radio image shows two extended radio lobes with a size $\sim$ 1.2 Mpc \citep{bassani2016}, plus a bright compact core at their center. Recent simultaneous multi-wavelength data 
have revealed that the nuclear emission is dominated by a jet pointing towards us, at low inclination angles \citep{lore2017}. One possible explanation to reconcile the large and small scale jets is a restarting activity scenario in which 
the source has initially produced two jets that are now observed as two extended old radio lobes. Lately, likely due to an intermittent activity episode, a new pair of jets have formed changing direction with respect to the old ones and pointing toward us, 
transforming this source from a radio galaxy to a blazar, a very exceptional case of restarting activity. 
Thus, although the host galaxy is very old, with stellar populations older than 10 Gyr, 
the nuclear radio core might be a very young source if we assume that the restarted activity occurred slowly, e.g., related to changes in the accretion, explaining the low nuclear bolometric luminosity (9.4$\times$10$^{43}$ erg\,s$^{-1}$, \citealt{lore2017}) compared to more powerful blazars 
($\sim$ 10$^{46}$ erg\,s${-1}$). The combination of the jet inclination angle and its low power
allows the detection of emission lines in the optical spectra, usually missed in blazars.
In this work, we show that PBC\,2333.9-2343 has an exceptional optical spectrum, where variable broad emission lines and broadened narrow emission lines are found, having a truly unique combination of BLR, outflow and jet interaction in the same galaxy. Here we present its exceptional characteristics at optical, UV, and X-ray frequencies.

The paper is organized as follows. In Section 2 we present the observations and data reduction, while the analysis and main results can be found in Section 3, which are discussed in Section 4. A summary of the main results is presented in Section 5. Throughout the text we have assumed H$_0$ = 70 km\,s$^{-1}$Mpc$^{-1}$, $\Omega_{\lambda}$ = 0.73 and $\Omega_M$ = 0.27, from which we have derived a luminosity distance of 206 Mpc, which corresponds to a scale of 0.92 kpc\,arcsec$^{-1}$.


\section{\label{reduction}Observations and data reduction}

The nucleus of PBC\,J2333.9$-$2343 was observed on 2009 September 18, on 2015 November 7, and on 2016 November 26 with the 2.12-m telescope of the San Pedro M\'artir Observatory (SPM, in M\'exico), equipped with a Boller \& Chivens spectrograph and a 2048$\times$2048 pixels E2V-4240 CCD, which was tuned to the 3800 \AA\ to 8000 \AA\ range (grating 300 l\,mm$^{-1}$), with a spectral dispersion of 4.5 \AA\ pixel$^{-1}$, corresponding to FWHM=10 \AA\, derived from the FWHM of different emission lines of the arc-lamp spectrum. 
A 2.5 $\arcsec$ slit was used oriented with the parallactic angle and the pixel spatial scale was 1.1 $\arcsec$\,pixel$^{-1}$ in 2009 and 0.6 $\arcsec$\,pixel$^{-1}$ in 2015 and 2016.
It was also observed once on 2016 November 16 with the 3.58-m New Technology Telescope (NTT, in Chile), using the ESO Faint Object Spectrograph and Camera \citep[EFOSC2,][]{buzzoni1984} which has a 2048$\times$2048 pixels CCD. The grism\#11 and 1$\arcsec$ slit was used oriented with the parallactic angle, corresponding to a resolution of 17.16 \AA\. The pixel spatial scale was 0.24 $\arcsec$\,pixel$^{-1}$ with a 2$\times$2 binning, and a spectral pixel scale of 2.04 \AA\ pixel$^{-1}$.
The spectrophotometric data reduction was carried out with the IRAF package\footnote{http://iraf.noao.edu/}, including bias and flat-field corrections, cosmic-ray removal, two-dimensional wavelength calibration, sky spectrum subtraction, and flux calibration.
It is worth noting that SPM spectra were acquired with a high airmass ($\sim$1.8), resulting into a slit loss and therefore cannot be used for absolute spectrophotometric measurements. However, given the reliability of line widths estimates, we use SPM spectra to study emission line variations, whereas the NTT spectrum is used to derive physical parameters. Note that SPM and NTT spectral results are not compared because of the different slit sizes which could introduce errors in the variability analysis.

We used all the public data of PBC\,J2333.9-2343 in the \emph{Swift} archive, eight times between 2010 and 2017. The data reduction of the \emph{Swift} X-ray Telescope \citep[XRT,][]{burrows2005} in the Photon Counting mode was performed by following standard routines described by the UK Swift Science Data Centre (UKSSDC)\footnote{http://www.swift.ac.uk/analysis/xrt/index.php} and using the software in HEASoft version 6.19. Calibrated event files were produced using the routine {\sc xrtpipeline}, accounting for bad pixels and effects of vignetting, and exposure maps were also created. Source and background spectra were extracted from circular regions with 30 $\arcsec$ and 80 $\arcsec$ radius. The {\sc xrtmkarf} task was used to create the corresponding ancillary response files. The response matrix files were obtained from the HEASARC CALibration DataBase. The spectra were grouped to have a minimum of 20 counts per bin using the {\sc grppha task}.

The Ultraviolet and Optical Telescope \citep[UVOT,][]{roming2005} observes with six primary photometric filters of V (centred at 5468 \AA), B (at 4392 \AA), U (at 3465 \AA), UVW1 (at 2600 \AA), UVM2 (at 2246 \AA) and UVW2 (at 1928 \AA). We notice that data are available in the six filters but not on all dates. The {\sc uvotsource} task was used to perform aperture photometry using a circular aperture radius of 5 $\arcsec$ again centred on the coordinates given by NED. The background region was selected as a free of sources circular region of 20 $\arcsec$ close to the nucleus.

\section{\label{analysis}Data analysis and results}

\subsection{ \label{opticalresults} Optical spectroscopy}

\begin{table*}
\begin{center}
\caption{\label{linefit}Line fitting of the optical spectra of PBC\,J2333.9-2343. \\
{\bf Notes.} For each spectrum, the signal to noise (S/N) of the continuum estimated in the 5650-5750 \AA\ region, where no emission or absorption lines are present, the reduced chi-square and degrees of freedom (d.o.f.) of the spectral fits for the [OIII]-H$\beta$/H$\alpha$-[NII] regions, respectively, and the seeing for each night are presented. From the fourth line on, for each line the centroid wavelength, $\lambda$, in \AA, shift respect to the NLR, $\Delta \lambda$, in \kms, line width, $\sigma$, in \kms, and fluxes, $F$, in units of 10$^{-15}$\,cm$^{-2}$\,s$^{-1}$ \AA$^{-1}$, are presented. The reported errors are those derived from the fit. The $\sigma$ values have been corrected for instrumental broadening.}
\begin{tabular}{lcccc|c} \hline \hline
Line & Parameter & SPM(2009) &  SPM(2015) &  SPM(2016) & ESO (2016) \\ 
(1) & (2) & (3) & (4) & (5) & (6)    \\ \hline
Spectrum & S/N & 39 & 43 & 39 & 66 \\
& $ \chi^2_r/d.o.f  $ & 0.9/1.0 & 0.8/1.0  & 1.3/1.2  & 1.1/1.0  \\
&  Seeing ($\arcsec$) & 2.5 & 2.6 & 2.5 & 1.2 \\
$[OIII]_{outflow}$ 
 & $\lambda$ & 5009.3$\pm$0.8 & 5007.4$\pm$0.3 & 5007.6$\pm$0.2 & 5001.8$\pm$3.1 \\
 & $\Delta \lambda$ & 144$\pm$49 & 36$\pm$19 & 30$\pm$13 & -347$\pm$187  \\
 & $\sigma$ & 850$\pm$60 & 880$\pm$30 & 803$\pm$24 & 1919$\pm$301 \\
 & $F$ & 13.8$\pm$2.0 & 19.4$\pm$1.1 & 13.1$\pm$0.6 & 8.3$\pm$1.9 \\
H$\alpha_{broad}$ 
 & $\lambda$ & 6622.9$\pm$6.2 & 6636.0$\pm$6.7 & 6654.9$\pm$3.0 & 6642.3$\pm$2.7 \\
 & $\Delta \lambda$ & 2738$\pm$283 & 3337$\pm$306 & 4201$\pm$137 & 3625$\pm$123 \\
 & $\sigma$ & 6340$\pm$283 & 5120$\pm$238 & 6235$\pm$183 & 5325$\pm$101 \\
 & $F$ & 45$\pm$3 & 66$\pm$5 & 115$\pm$5 & 80$\pm$2 \\
H$\beta_{broad}$ 
 & $\lambda$ & - & - & 4935.0$\pm$6.3 & 4879.7$\pm$21.9 \\
 & $\Delta \lambda$ & - & - & 4567$\pm$389 & 1154$\pm$1352 \\
 & $\sigma$ & -& - & 4005$\pm$259 & 5110$\pm$1062 \\
 & $F$ & - & - & 9$\pm$1 &  13$\pm$4  \\
\hline                      
\end{tabular} 
\end{center}
\end{table*}

\begin{figure*}
\centering
\includegraphics[width=0.4\textwidth]{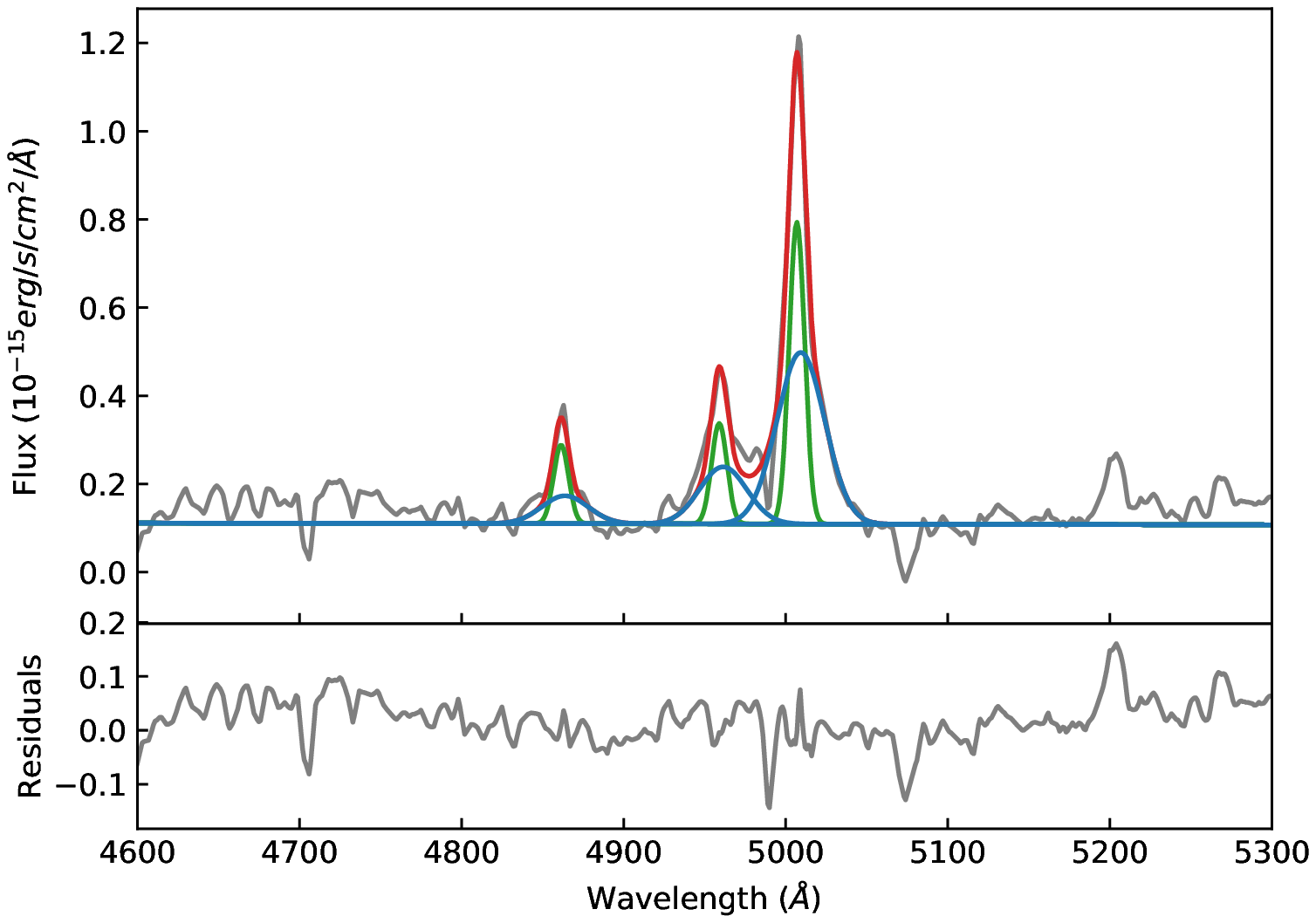}
\includegraphics[width=0.4\textwidth]{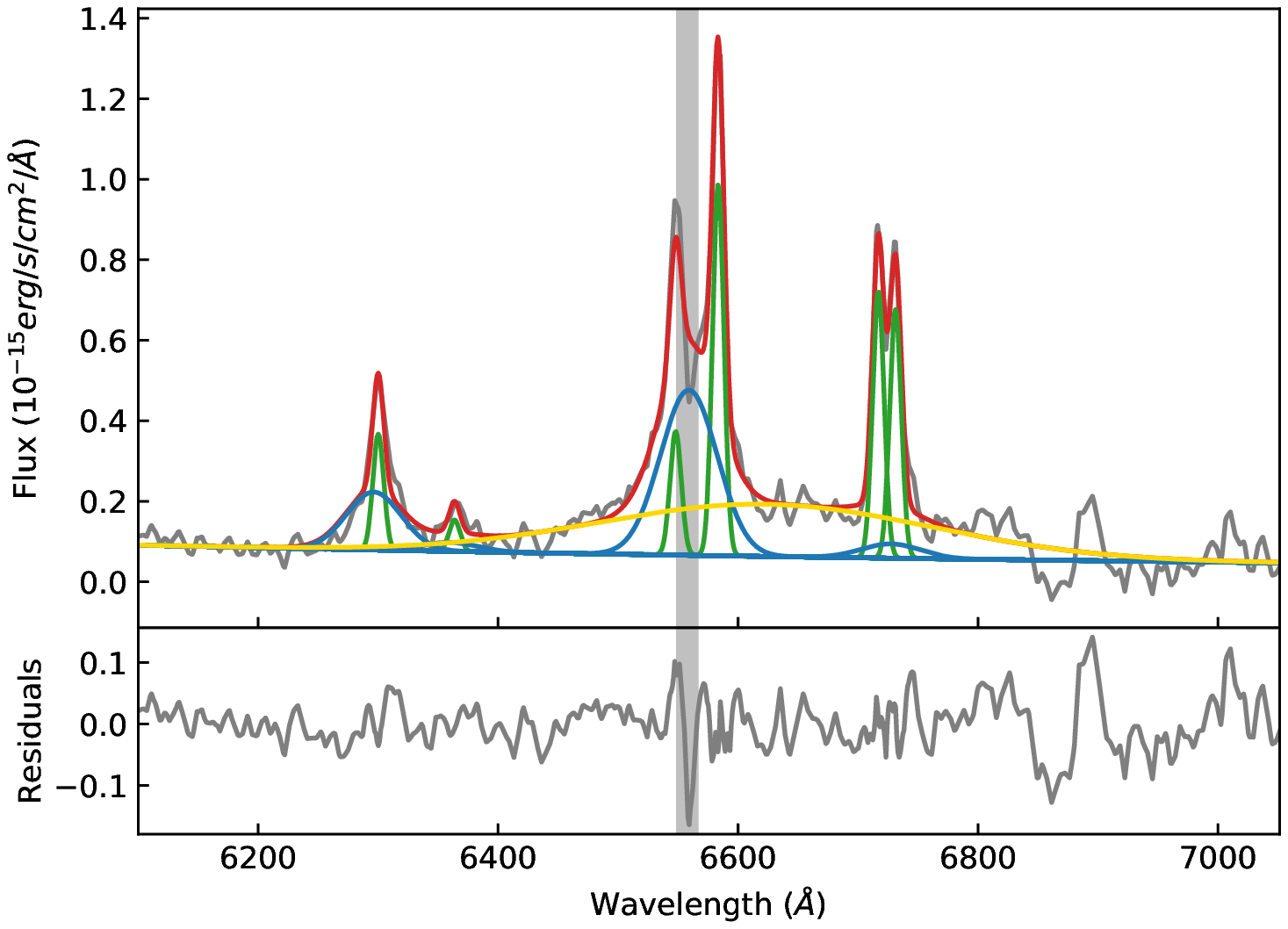}

\includegraphics[width=0.4\textwidth]{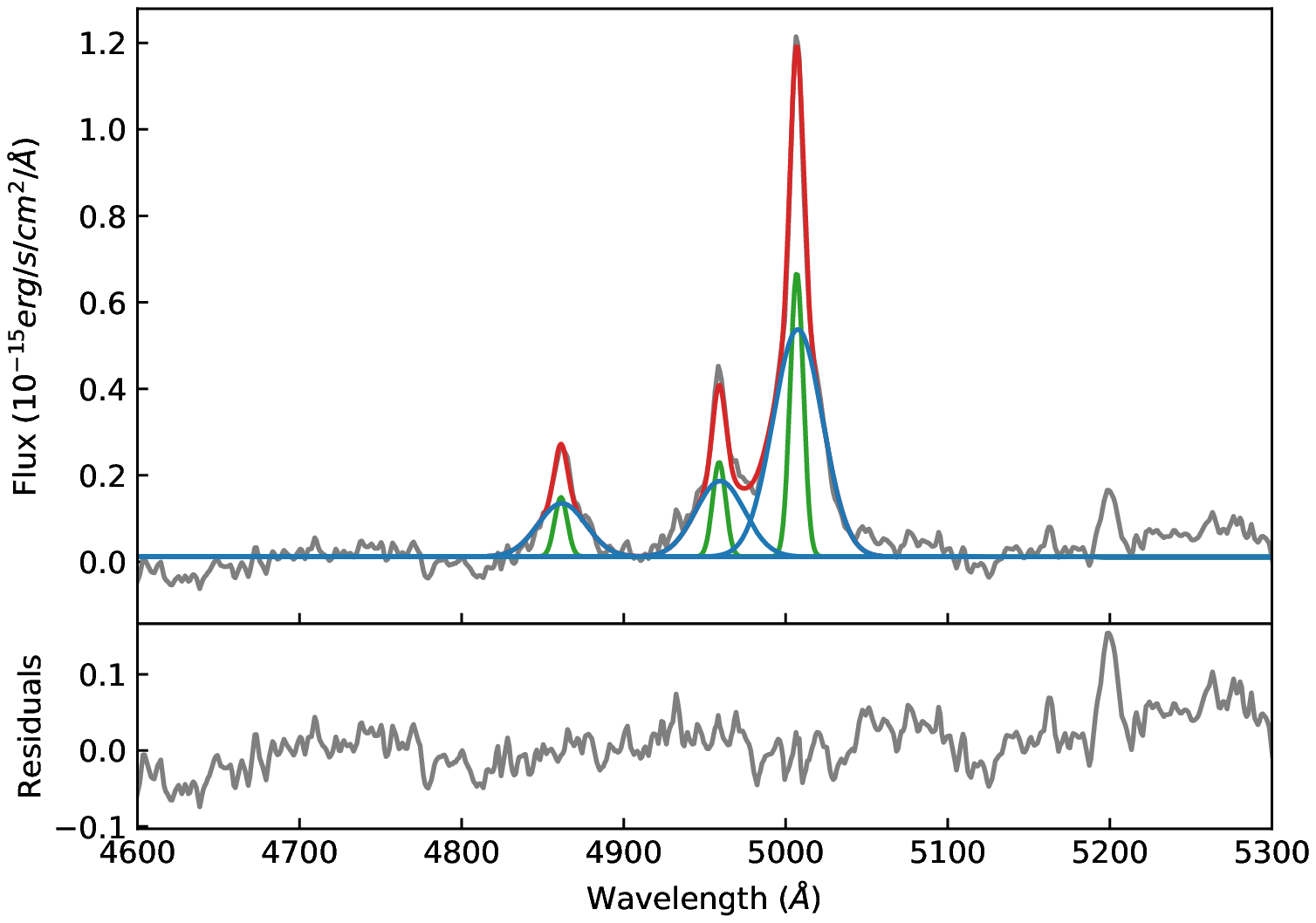}
\includegraphics[width=0.4\textwidth]{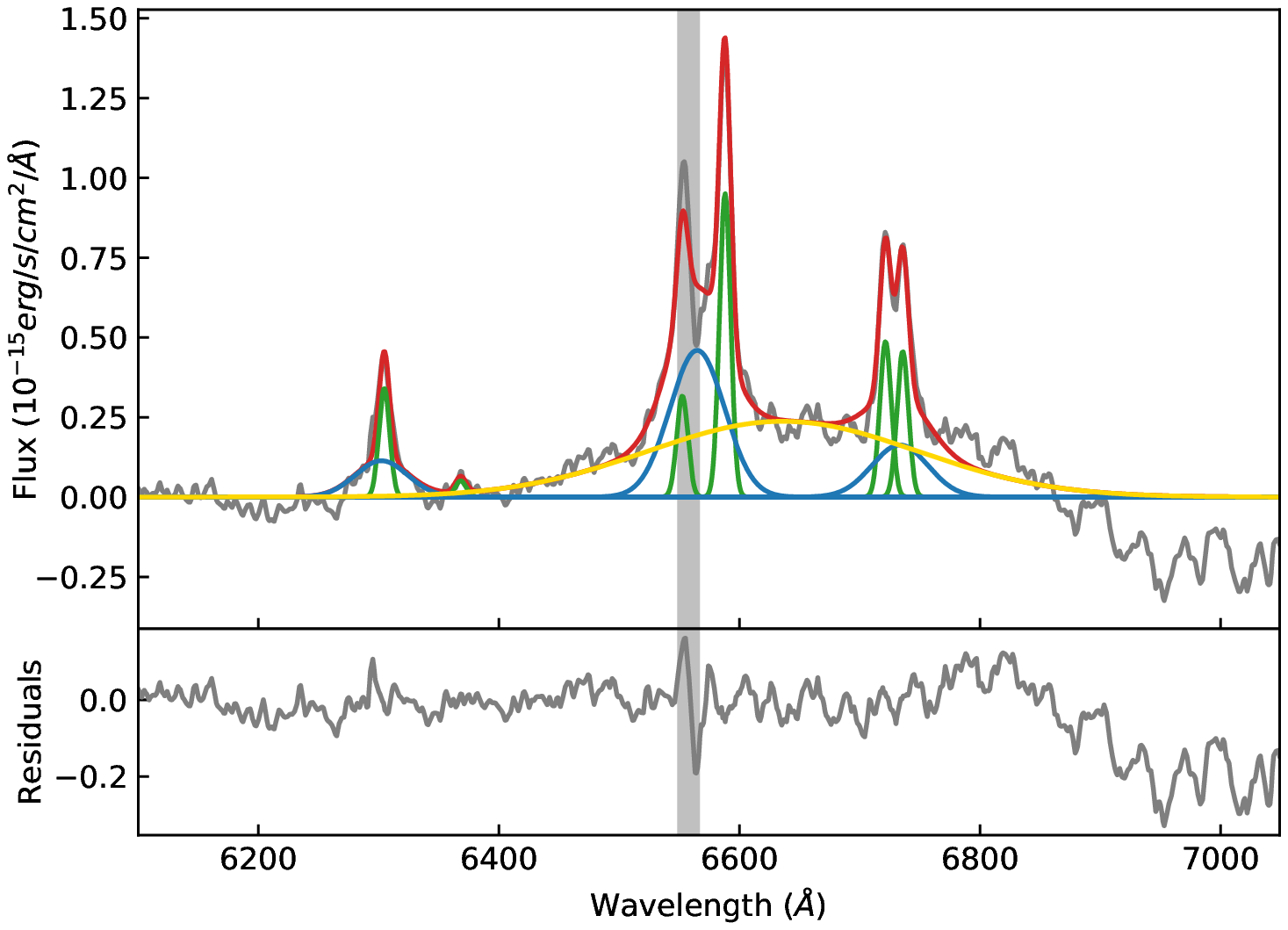}

\includegraphics[width=0.4\textwidth]{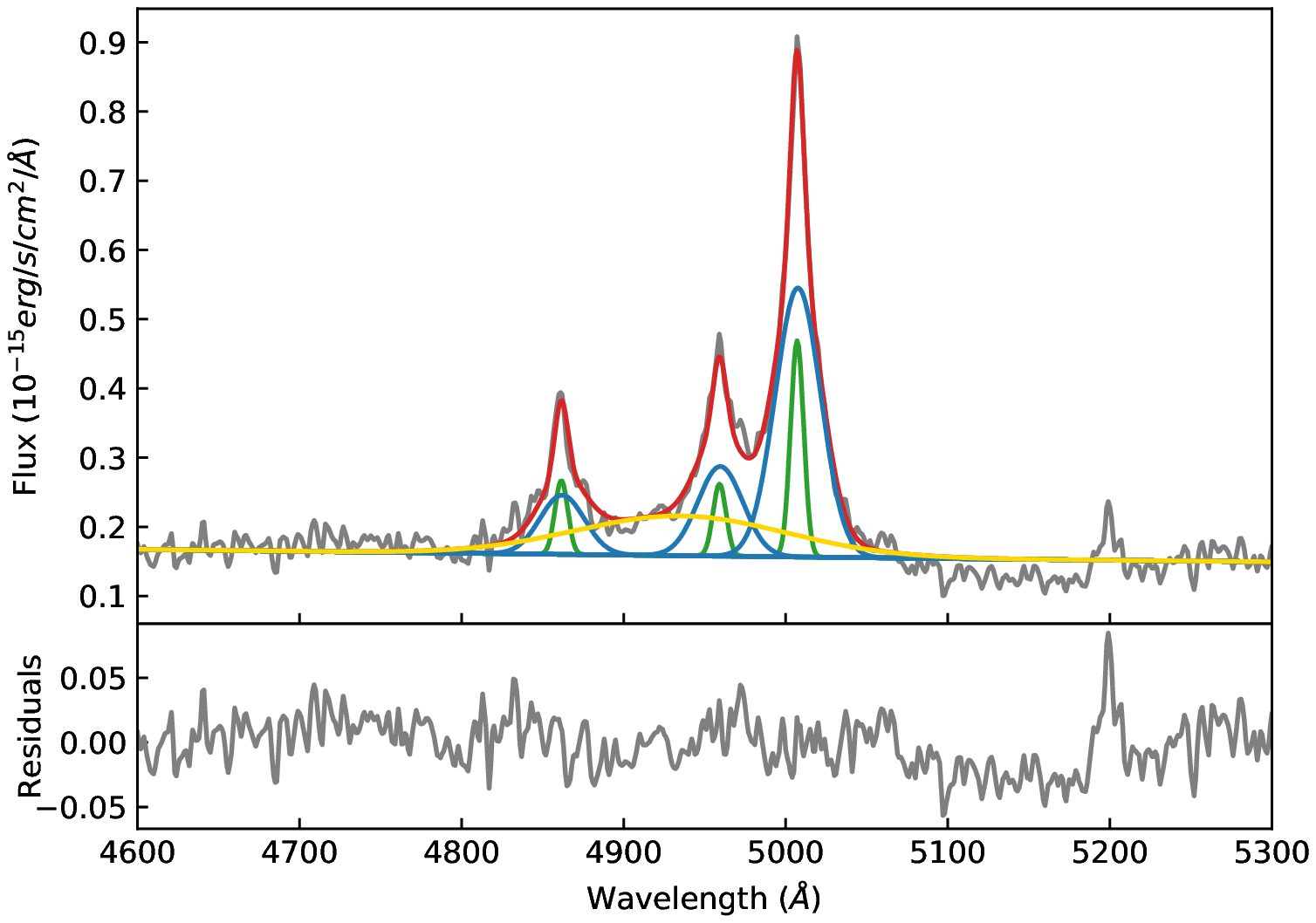}
\includegraphics[width=0.4\textwidth]{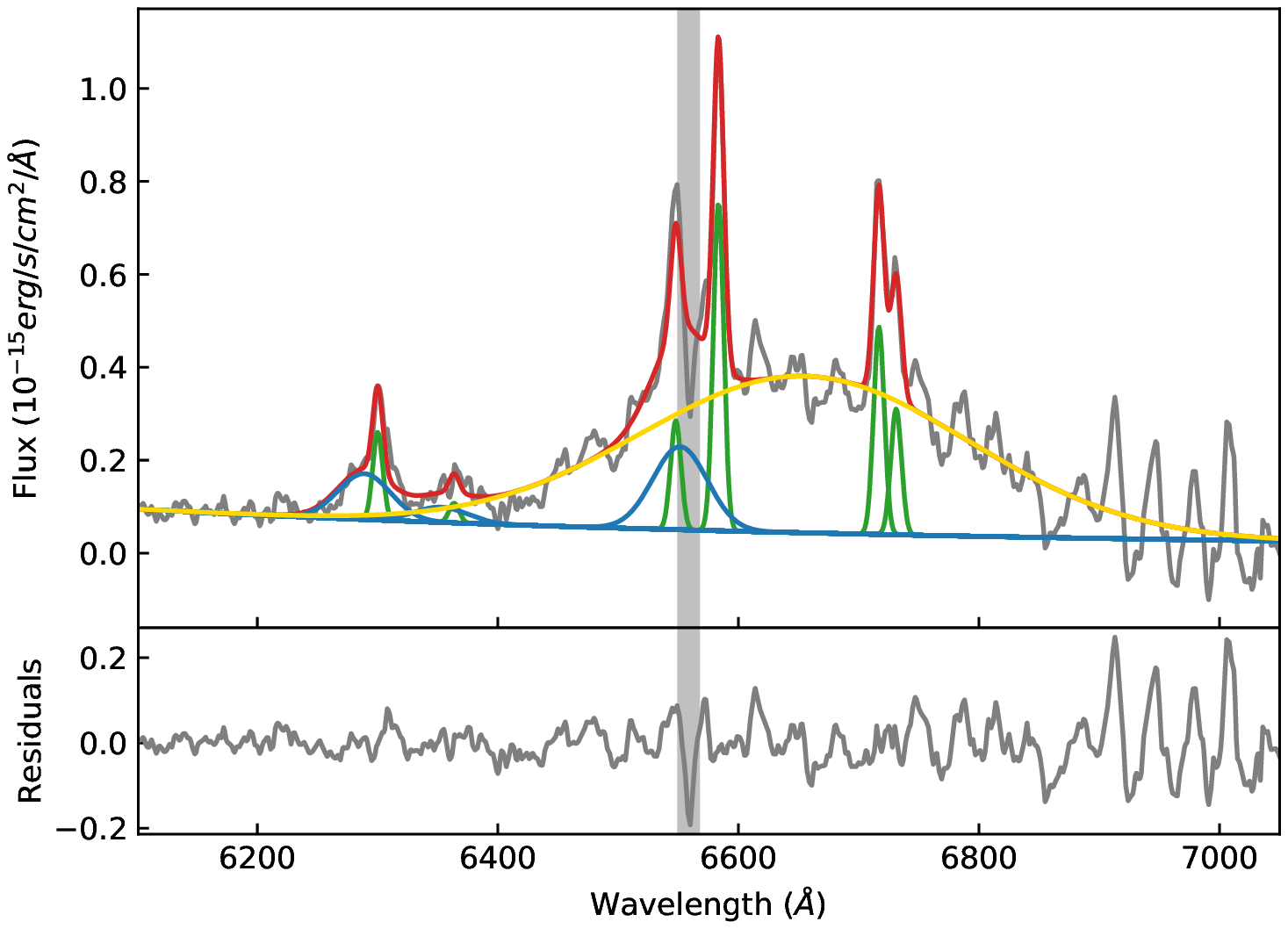}

\includegraphics[width=0.4\textwidth]{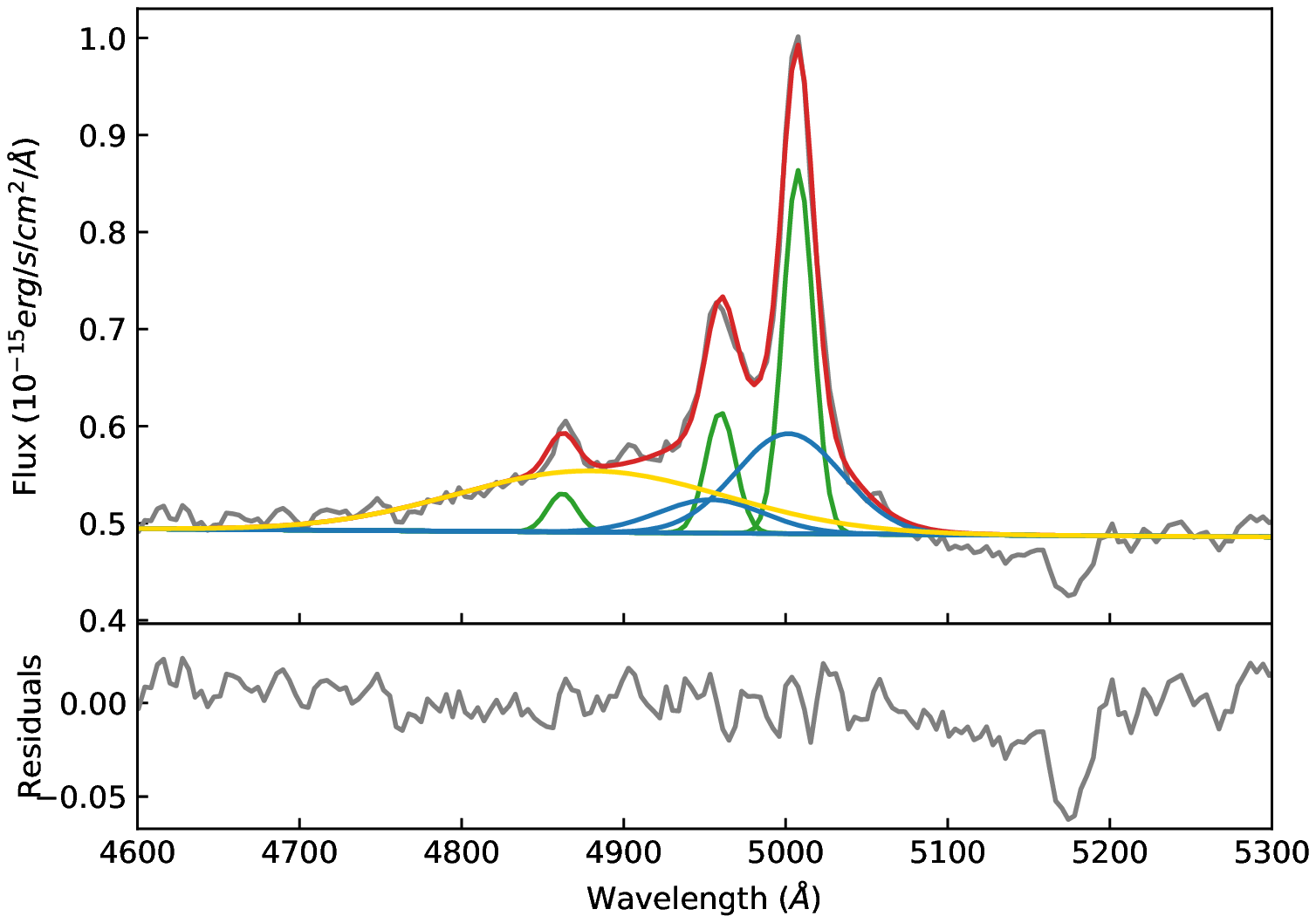}x
\includegraphics[width=0.4\textwidth]{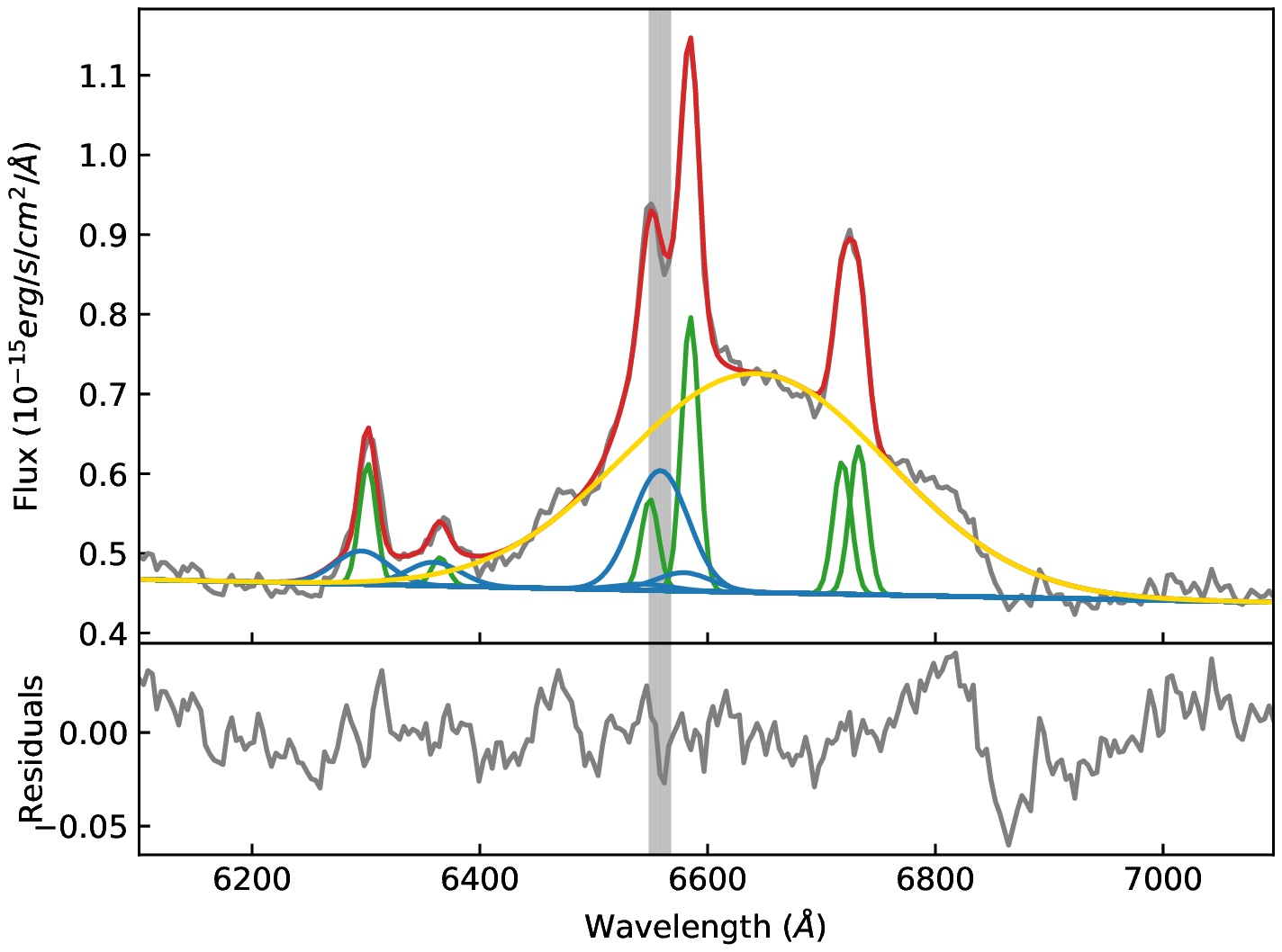}
\caption{\label{optical} [OIII]-H${\beta}$ and H${\alpha}$ regions of SPM and ESO optical spectrum and its residuals. From up to bottom: 2009, 2015, 2016 (SPM) and 2016 (ESO). The green lines represent the narrow components, blue the outflow, yellow the broad lines from the BLR, and red is the total model. The grey area represents the wavelength zone affected by the $O_2$ telluric line.}
\end{figure*}

We applied the STARLIGHT stellar population synthesis code \citep{cid2005} for the recovery of the shape of the stellar continuum to all the spectra. 
We performed the spectral synthesis using 150 single stellar populations from the evolutionary synthesis models of \cite{bruzual2003} and six different power-law slopes to simulate the AGN component to the optical continuum emission. The extinction law of \cite{cardelli1989} with Rv = 3.1 was adopted. The base components comprise 25 ages between 1 Myr and 18 Gyr, and six metallicities, from Z=0.005 $Z_{\odot}$ to 2.5 $Z_{\odot}$. 
The regions affected by atmospheric absorption and emission lines were masked and excluded from the fit. From the NTT data, we estimated the stellar velocity dispersion, $\sigma_{\star}$=269$\pm$4 \kms.

We then developed a model based on the scipy.optimize package in Python, using the curve fit routine, in order to perform the spectral analysis on the stellar population-subtracted spectra.
The [OIII]-H${\beta}$ and H${\alpha}$-[NII] regions were fitted separately, using the following models:

--{\it  [OIII]-H${\beta}$:} (i) a Gaussian profile for fitting each narrow component of the [OIII]$\lambda$4959,5007 \AA\ doublet and the H${\beta}$ emission lines, associated with the NLR; (ii)  the same number of Gaussians as in (i) for fitting the possible broad components associated with the outflowing gas; (iii) one Gaussian profile for fitting the BLR component of the H${\beta}$ emission.

 --{\it  H${\alpha}$-[NII]:} (i) a Gaussian profile for fitting each narrow component of the [OI], [NII] and [SII] doublets and of the H${\alpha}$ emission line, associated with the NLR; (ii) the same number of Gaussians as in (i) for fitting the possible broad components associated with the outflowing gas; (iii) one Gaussian profile for fitting the BLR component of the H${\alpha}$ emission.\footnote{We notice that the H${\alpha}$ broad component associated with the BLR was fitted in all spectra. The BLR H${\beta}$ emission was only added in the 2016 spectra.}

We fitted the spectra by fixing the relative wavelengths and intensities of the lines according to their atomic parameters, and assuming equal broadening for the lines of the same kinematic component. We limited the width of the narrow component to be larger than the instrumental one. 

The normalizations between the [OIII] doublet was fixed equal to 1:3. The same ratio was applied to the [NII] doublet. 
Moreover, the intrinsic velocity dispersion of each emission line was forced to be the same in each fitted region.
Same rules were applied for the broad components associated with the outflow. The measurements were corrected for instrumental broadening.
Best-fit results for [OIII]$\lambda$5007, H${\beta}$ and H${\alpha}$ are presented in Table \ref{linefit} and Figure \ref{optical}. 

The main results obtained from the optical spectral fitting are the following:
\\
-- {\it Broad Balmer emission line variations}: by comparing the three SPM spectra, the strongest variations are found in the broad H$\alpha$ component (Fig. \ref{optical}). 
Specifically, it shows a flux brightening of 61$\pm$4 per cent between 2009 and 2016. 
Moreover, the coherence between the 2016 SPM and NTT spectra, taken in very different observing conditions and with very different spectral set up, support the remarkable increase in the relative contribution of the broad H$\alpha$ component.
In the 2016 NTT and SPM spectra we also detected a broad H$\beta$ component, which was not present in the 2009 and 2015 SPM spectra. 
This implies that the optical classification of this source has changed from type 1.9 to type 1.8 in one year. 
\\
--  {\it Broad Balmer emission line shifts}: we have measured a line centroid of the H${\beta}$ and H${\alpha}$ redshifted by $>$1000 \kms\ with respect to the systemic velocity, which is estimated from the narrow component of the [OIII] emission line.
Unfortunately, a telluric line affects the H${\alpha}$ wavelength region (Fig. \ref{optical}), preventing us from deriving the correct line profile and thus from confirming the exact value of the shift. Note that the shift is within the errors in the NTT spectrum, thus this result still needs to be confirmed.
\\ 
--  {\it Broad component of the narrow emission lines}: a broad blueshifted component is confirmed in high ([OIII] doublet) and low ionization lines (H$\beta$, [OI]$\lambda$6300) in all spectra. Here we study the broad component of the [OIII] emission line.

Finally, we have used independent methods to provide a black hole mass estimate. 
Using FWHM(H$\beta$) and the continuum at 5100 \AA\ following \cite{bongiorno2014} and their error estimation, we measure $log(M_{BH}^{H{\beta}}/M_{\odot})=8.4 \pm 1.4$, where the luminosity at 5100 \AA\ was estimated from the AGN component in the STARLIGHT fit (L5100\AA\ = (1.3$\pm$0.4)$\times$10$^{43}$ erg\,s$^{-1}$). We notice that if the jet partially ionizes the BLR (see Sect. 4.1), the motions of the BLR would be at least partially non gravitational and this would imply an incorrect value of $M_{BH}$.
Using the $\sigma_{\star}$ derived from the NTT spectrum and the relation in \cite{gultekin2009}, we obtain $log(M_{BH}^{\sigma_{\star}}/M_{\odot})=8.67 \pm 0.13$.  
In addition, using the K-band from the 2MASS catalogue as reported in NED and following \cite{marconi2003}, we found $log(M_{BH}/M_{\odot})=8.4$.
These measurements are consistent within uncertainties and in agreement with previous ones in \cite{lore2017}.

\subsection{\label{xrayresults}X-ray spectroscopy, UV and optical photometry}

\begin{figure}
\includegraphics[width=0.4\textwidth]{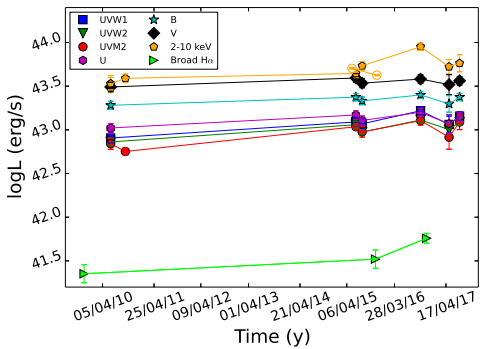}
\caption{\label{uvx} \emph{Swift}/XRT (2--10 keV energy band), \emph{Swift}/UVOT (UV/optical), and broad H$\alpha$(SPM) intrinsic luminosities of PBC\,J2333.9-2343. The orange empty circles correspond to the 2--10 keV luminosities from \emph{XMM-Newton} presented in Hern\'andez-Garc\'ia et al. (2017b).}
\end{figure}

The X-ray data analysis was performed using XSPEC v. 12.9.0. We assume a Galactic absorption of N$_{Gal}$ = 1.63$\times$10$^{20}$ cm$^{-2}$  \citep{dickey1990}.
Following the results obtained in \cite{lore2017}, we fitted the \emph{Swift}/XRT spectra of with a single power-law model.
All the spectra were fitted simultaneously to avoid degenerations between the spectral parameters, and because this process allows to search for X-ray spectral variability \citep{lore2013}. We found that linking all the parameters to each other does not result in a good spectral fit ($\chi^2$/do.f.=3.18), indicating presence of variability. We let the spectral index, $\Gamma$, and normalization of the power law, vary one-by-one and by pairs in the model. We found that the model where the normalization varied and $\Gamma$=1.65$_{-0.04}^{+0.05}$ remained constant showed the greatest improvement ($\chi^2$/do.f.=1.09) respect to the model where $\Gamma$ is varying ($\chi^2$/do.f.=2.58). The X-ray fluxes derived from the fit are presented in Col. 4 of Table \ref{observations}.
These values are in agreement with those obtained with \emph{XMM-Newton} in 2015 (two observations separated by six months, see Fig. \ref{uvx}), where a small drop of 16 per cent was detected \citep{lore2017}. 
We also let the two parameters vary together, but the improvement is not statistically significant ($\chi^2$/do.f.=1.08). 
UVOT data showed UV and optical variations between the observations above the 3$\sigma$ confidence level.

Overall, our results show that PBC\,J2333.9-2343 is variable at all the observed frequencies on timescales between months and years (Fig. \ref{uvx}).
The largest flux variations are obtained between 2010 and 2016, with a brightening of 62$\pm$6 per cent (2-10 keV), 19$\pm$8 per cent (V), 24$\pm$6 per cent (B), 34$\pm$5 per cent (U), 51$\pm$5 per cent (UVW1), 56$\pm$4 per cent (UVM2), and 44$\pm$5 per cent (UVW2).
Moreover, the optical continuum is bluer in 2016 than in 2009, in agreement with the UV variations.

\begin{table*}
\begin{center}
\caption{\label{observations}\emph{Swift}/XRT observations of PBC\,J2333.9-2343. \\
{\bf Notes.}  (Col. 1) ObsID., (Col. 2) date, (Col. 3)  exposure time in ksec, (Col. 4) X-ray flux in units of 10$^{-11}$ erg\,s$^{-1}$cm$^{-2}$ from the XRT when fitting a power-law model, and (Cols. 5-10) UV and optical fluxes in units of 10$^{-15}$ erg\,s$^{-1}$cm$^{-2}$ from the UVOT.}
\begin{tabular}{cccccccccc} \hline \hline
ObsID. & Date & Exposure & 2-10 keV & UVW2 & UVM2 & UVW1 & U & B & V \\ 
(1) & (2) & (3) & (4) & (5) & (6) & (7) & (8) & (9) & (10) \\ \hline
00031731001 & 2010-06-05 & 2  & 0.64$_{-0.06}^{+0.07}$ & 0.82$\pm$0.04 & 0.67$\pm$0.04 & 0.67$\pm$0.04 & 0.65$\pm$0.03 & 0.95$\pm$0.04 & 1.23$\pm$0.07\\
00031731002 & 2010-06-05 & 4  & 0.65$_{-0.05}^{+0.05}$ &  	-     &  	-     &  	 -    & 0.66$\pm$0.01 &  	-     &  	-    \\
00041128001 & 2010-09-23 & 9  & 0.74$_{-0.03}^{+0.04}$ &  	-     & 0.55$\pm$0.02 &  	-     &      - 	     &  	-     &  	-    \\
00081308001 & 2015-06-13 & 7  & 0.84$_{-0.04}^{+0.05}$ & 1.29$\pm$0.05 & 1.05$\pm$0.04 & 1.03$\pm$0.05 & 0.93$\pm$0.04 & 1.17$\pm$0.04 & 1.56$\pm$0.07\\
00081308002 & 2015-07-30 & 6  & 1.03$_{-0.04}^{+0.05}$ & 1.06$\pm$0.03 & 0.92$\pm$0.05 & 0.98$\pm$0.05 & 0.80$\pm$0.04 & 1.06$\pm$0.04 & 1.36$\pm$0.07\\
00031731003 & 2016-10-09 & 6  & 1.71$_{-0.08}^{+0.07}$ & 1.46$\pm$0.06 & 1.23$\pm$0.06 & 1.38$\pm$0.06 & 1.00$\pm$0.04 & 1.24$\pm$0.05 & 1.51$\pm$0.07\\
00031731004 & 2017-05-10 & 3  & 1.01$_{-0.08}^{+0.08}$ & 1.13$\pm$0.08 & 0.80$\pm$0.11 & 0.96$\pm$0.09 & 0.74$\pm$0.07 & 0.98$\pm$0.09 & 1.31$\pm$0.01 \\
00031731005 & 2017-07-28 & 5  & 1.11$_{-0.12}^{+0.13}$ & 1.38$\pm$0.06 & 1.19$\pm$0.10 & 1.20$\pm$0.06 & 0.88$\pm$0.04 & 1.17$\pm$0.04 & 1.44$\pm$0.07\\
\hline                      
\end{tabular} 
\end{center}
\end{table*}

\section{\label{discusion}Discussion}

\subsection{Continuum and line variability}

In non-jetted AGN, the origin of emission line variations is ascribed to variations from the photo-ionizing continuum \citep{peterson1988}, or to extinction caused by clouds intersecting the line of sight of the observer \citep[e.g.,][]{lore2017seyfert}. 
Blazars are usually characterized by absent or very weak optical emission lines, because of the strong and highly variable over-imposed continuum emission \citep{netzer2013}. 
In the case of \pks\, the low power jet and favorable jet inclination angle allow us to detect strong emission lines and their variability. It is unlikely that obscuration is the origin of the variability in \pks, as the source is not absorbed at 
X-rays and does not show spectral shape variations.
The strong variability observed in UV and X-rays favors the hypothesis that the emission line fluxes are responding to the continuum variability. 

Fig. \ref{uvx} shows the simultaneous X-ray and UV/optical variations observed by \emph{Swift}, suggesting that the regions responsible for the X-ray and UV emissions are coupled over long timescales.
The flux of the broad H$\alpha$ increased by 61$\pm$4 per cent between 2009 and 2016, similarly the X-ray variation is of 62$\pm$6 per cent between 2010 and 2016. Also the flux of the broad component of the [OIII] line 
has increased by 29$\pm$14 per cent between 2009 and 2015. This variation is however detected at a 2$\sigma$ confidence level and it deserves further investigation. 
Future monitoring of this source with high quality spectroscopy will shed light on these spectral variations.

According to the results of the spectral energy distribution fitting, the X-rays in this source are produced by the jet \citep{lore2017}. The evidence that in our observations the broad H$\alpha$ flux increases in time as well as the X-ray flux,
suggests that the jet could be, at least in part, responsible for the ionization of the BLR and, if variations will be confirmed in the outflowing component, also in the NLR.

The interaction of BLR clouds with a jet has already been suggested for AGN. For instance, line variations in Mg II$\lambda$2800 related with a flare-like event have been found in the blazar 3C\,454.3 \citep[where also a change in the jet direction has been observed,][]{raiteri2008}, indicating that the jet powered the BLR during intense outbursts \citep{leontavares2013}. In the case of 3C~277.3, a strong relationship between emission-line gas and the radio jet has been found by \cite{vanbreugel1985}, who have suggested that the collision between the jet and the clouds produces flaring radio emission and the ionizing synchrotron continuum (optical/infrared), and deflects and entrains the jet material. 
The action of the jet to ionize the surrounding medium could be exerted in two ways. One one hand, it could be collisionally excited as a result of heating \citep{paltani2003, punsly2013}. On the other hand, the hot shocked gas could  emit intense UV and soft X-ray radiation, photoionizing the gas ahead of the shock \citep{allen2008}. New data are however needed in order to determine the correct profile of each line and their shifts and to study the effect of the jet.

\subsection{Outflow}

The [OIII] emission line is produced in low density ambient, e.g., at large scales in the NLR. 
In a large number of AGN, the spectral profile of the [OIII]$\lambda$5007 can be highly asymmetric, showing a broad blue-shifted profile,
typically associated with outflowing ionized gas at kpc-scale \citep{fiore2017,perna2017}.  
Broad [OIII] components are observed in all the optical spectra presented here, however, because of the higher reliability in the flux calibration of the lines, we focus on the NTT spectrum to derive the physical parameters of the outflow.
The measured [OIII] component is quite broad ($\sigma$=1919$\pm$301 \kms), with a modest blueshift (347$\pm$187 \kms) and a maximum velocity of $| \Delta v| + 2 \sigma$=4230$\pm$660 \kms , representing the outflow velocity 
(assumed constant with radius and spherically symmetric) along the line of sight. 
The low shift and the large FWHM could be due to: i) a gas highly perturbed \citep{villarmartin2011}, possibly by the jet itself; ii) a particular geometry for which we can see both the redshifted and the blueshifted components of the outflow, 
similarly to the quasar SDSS\,J1201+1206 that shows a quasi symmetric line \citep{bischetti2017}; or a combination of (i) and (ii). 

The derived mass outflow rate is $\dot{M}$=(2.3$^{+5.8}_{-0.4}$)$ M_{\odot}$,yr$^ {-1}$, and the kinetic energy is $E_k$=(1.3$^{+3.2}_{-0.2}$)$\times$10$^{43}$ erg\,s$^{-1}$, obtained using equations 5 and 6 from \cite{bischetti2017}, and following their error estimate. We assumed C$\sim$1, [O/H]$\sim$0 (solar metallicity), an electron density n$_e$=200 cm$^{-3}$ 
(an estimate of the density from the [SII] was not achievable because of the blending in both the [SII] and [OIII]$\lambda$4363 lines), and a radius of the emitting broad [OIII] region of 0.46 kpc, estimated from the size of the aperture used to extract the NTT spectrum, which represents an upper limit \citep{villarmartin2016}.
We caution that the derived values are rough estimates because of the several assumptions \citep[see details in][]{kakkad2016}.

We have compared the above values with those obtained for different kind of winds 
 as reported in \cite{fiore2017}. They compiled a sample of 
94 AGN with detected outflows at sub-pc to kpc spatial scales and investigated scaling relations between AGN, host galaxy and outflows properties. 
\pks\ is characterized by a relatively low bolometric luminosity 
with respect to other AGN producing winds. Interestingly, the $E_k$ and $\dot{M}$ values place \pks\ at the low end 
distribution of ionised winds (see Figure 1, in \citealt{fiore2017}). On the other hand, the measured maximum velocity is relatively high given its bolometric and broad [OIII] luminosity (4.2$\pm$1.0$\times$10$^{40}$ erg\,$s^{-1}$).
However, the main assumption in \cite{fiore2017} is that winds are radiation-driven, while in the case of \pks\ the presence of a blazar-like jet should be taken into account as possible source of radiation and/or shocks production. 
The extreme kinematics of the broad component in [OIII] compared to the stellar kinematics and its low bolometric luminosity indeed suggest that the outflow is induced by the interaction between the jet and the ambient gas. This is in agreement
with the fact that, even in radio quiet objects, the most extreme outflows are often related with objects presenting modest level of jet activity \citep{whittle1992, tadhunter2000,mullaney2013,villarmartin2014,tadhunter2014,giroletti2017}. 

Outflows are also observed in radio loud sources \citep[e.g.,][]{emonts2005,morganti2013,mahony2016,couto2017}.  
In particular, the co-existence of a mildly relativistic wind detected at X-rays with velocities $\sim 0.1c$ and a jet in a radio galaxy \citep{tombesi2012}, indicates a transverse stratification of a
global flow and the possibility that the X-ray wind could provide additional support for the initial jet collimation \citep{fukumura2014}. 
The radio activity is very likely influencing the energetics and thermodynamics of the emission-line gas during the period that the jet is expanding through the interstellar medium \citep[e.g.,][]{best2000}. 
Recently, \cite{couto2017} used integral field spectroscopy to study the kinematics of the radio galaxy 3C\,33, proposing that shocks in the jet lead to a lateral expansion of the gas, producing an outflow, which is responsible for shock ionization of the gas. Another example is 3C\,293, a double-double radio galaxy that shows a jet-driven outflow of ionised gas \citep{emonts2005,mahony2016}. 
Theoretically, current models of interactions between the jet and the interstellar medium predict that these kind of outflows are produced by shocks that ionise the surrounding medium \citep{wagner2012}.

\subsection{Caveats and limitations of the analysis}

The SPM spectra presented in this paper are affected by some source of uncertainties and thus we caution the reader on some aspects of the presented results. 
The spectra were obtained with a high airmass and a telluric line affects the H$\alpha$ region, preventing the determination of the line centroid of its narrow component, 
and consequently on the [NII]/H$\alpha$ and [SII]/H$\alpha$ line ratios. 
We notice that by fitting the narrow and outflow components in the red lines our aim is to constrain as
accurately as possible their combined contribution to the total [NII]-H$\alpha$ blend
and to estimate how different fits could affect the parameters of the underlying
broad H$\alpha$ component. For this reason, the fluxes and relative line ratios of the
narrow and outflow components in the red lines are not expected to be accurate.
We confirm that the large associated uncertainties affecting them do not affect the
conclusions regarding the broad underlying H$\alpha$, whose errors also account for
the diversity of fits.
Despite the above, the estimates on line widths are reliable and statistically significant, for instance the variation of the width of the broad H$\alpha$ component is detected at a 14$\sigma$ confidence level. Variations in the broad lines might also be related to changes in the continuum level, i.e., intrinsic to the nuclear source.
However, from the optical photometry obtained with \emph{Swift}/UVOT data, variations in the continuum are at the level of $\sim$20 per cent, with much smaller amplitude than those observed in the broad H$\alpha$ component, thus supporting the variations in the BLR.

A huge shift in the velocity of the broad H$\alpha$ larger than 1000 \kms\ might be present. Since the narrow H$\alpha$ component is suppressed due to the telluric line, we cannot confirm the exact shift of the line. Space spectroscopic observations are needed to measure the exact shift of the broad H$\alpha$ component.

\section{Conclusions}

In the present work we present multi epoch optical (SPM and NTT), and UV and X-ray (\emph{Swift}) data of the nucleus in the peculiar source PBC\,J2333.9-2343, a blazar candidate. These data reveal:

\begin{itemize}
\item A strong and variable broad component of H${\alpha}$, with a flux variation of 61 per cent between 2009 and 2016.
\item A broad component in H${\beta}$, observed in the 2016 spectra, but not present in 2009 nor in 2015, changing the optical classification from a type 1.9 to a 1.8 Seyfert.
\item A broadened, blueshifted component is confirmed in high ([OIII] doublet) and low ionization lines (H$\beta$, [OI]$\lambda$6300).
\item An X-ray variable flux (amplitude of $\sim$ 62 per cent) between 2010 and 2016.
\item An optical and UV variable continuum ($\sim$20 per cent and $\sim$50 per cent, respectively).
\end{itemize}

The spectral X-ray and broad H$\alpha$ component variations are of the same order, they occur during the same timescale and they show the same trend. Since X-rays are interpreted as emission from the jet, we propose that the BLR is at least in part responsible for the ionization of the AGN. Moreover, the extreme kinematics of the broadened component in [OIII] and its low bolometric luminosity suggest that the outflow is induced by the interaction between the jet and the ambient gas.

\section*{Acknowledgments}

We acknowledge  the referee,  M. Villar Mart\'in,  for her comments  and  suggestions  that  helped  to  improve  the  paper.
This work made use  of  data  supplied  by  the  UK \emph{Swift} Science  Data  Centre  at the  University  of  Leicester, the NASA/IPAC
extragalactic database (NED), the STARLIGHT code, and the IRAF software. 
Based upon observations carried out at the Observatorio Astron\'omico Nacional on the Sierra San Pedro M\'artir (OAN-SPM), Baja California, M\'exico.
LHG and FP acknowledge the ASI/INAF agreement number 2013-023-R1, LHG partial support from FONDECYT through grant 3170527, MP from ESSTI under the MoST, and from MINECO through research projects
AYA2013-42227-P and AYA2016-76682-C3-1-P (AEI/FEDER, UE), SC from the Spanish grant AYA2013-42227-P, VC by CONACyT research grant 280789, EFJA from the Collaborative Research Center 956, subproject A1, funded by DFG, and GV from the DFG Cluster of Excellence ‘Origin and Structure of the Universe’ (www.universe-cluster.de).

\bibliographystyle{mn2e}
\bibliography{000referencias}

\end{document}